\documentclass[aps,prl,twocolumn,showpacs,byrevtex]{revtex4}

\let\oldhat\hat
\renewcommand{\hat}[1]{\oldhat{\mathbf{#1}}}

\usepackage{times,mathptm}
\usepackage[dvips]{graphicx,color}
\usepackage{titlesec}

\begin{document}

\title{Underlying mechanism of charge transfer in Li-doped MgH$_{16}$ at high pressure}
\author{Chongze Wang, Seho Yi, and Jun-Hyung Cho$^{*}$}
\affiliation{
Department of Physics, Research Institute for Natural Science, and HYU-HPSTAR-CIS High Pressure Research Center, Hanyang
University, 222 Wangsimni-ro, Seongdong-Ku, Seoul 04763, Republic of Korea}
\date{\today}

\begin{abstract}
A lithium-doped magnesium hydride Li$_2$MgH$_{16}$ was recently reported [Y. Sun $et$ $al$., Phys. Rev. Lett. {\bf 123}, 097001 (2019)] to exhibit the highest ever predicted superconducting transition temperature $T_{\rm c}$ under high pressure. Based on first-principles density-functional theory calculations, we reveal that the Li dopants locating in the pyroclore lattice sites give rise to the excess electrons distributed in interstitial regions. Such loosely bound anionic electrons are easily captured to stabilize a clathrate structure consisting of H cages. This addition of anionic electrons to H cages enhances the H-derived electronic density of states at the Fermi level, thereby leading to a high-$T_{\rm c}$ superconductivity. We thus propose that the electride nature of Li dopants is an essential ingredient in the charge transfer between Li dopants and H atoms. Our findings offer a deeper understanding of the underlying mechanism of charge transfer in Li$_2$MgH$_{16}$ at high pressure.
\end{abstract}

\maketitle

\section{I. INTRODUCTION}

The realization of room-temperature superconductivity (SC) is one of the most challenging and very long standing issues in condensed matter physics~\cite{Hydride-Rev2020-Eremets}. For this issue, metallic hydrogen has been proposed~\cite{Ashcroft} as an ideal system exhibiting the conventional Bardeen-Cooper-Schrieffer (BCS) type SC~\cite{BCS}. The lightest atomic mass in metallic hydrogen gives rise to high vibrational frequencies, which lead to achieving high $T_{\rm c}$ with strong electron-phonon coupling. However, it is very challenging to synthesize metallic hydrogen under extremely high pressures over ${\sim}$400 GPa~\cite{MetalicH-Rev.Mod.Phys2012,MetalicH-PRL2015,MetalicH-Science2017,hydrogen2020} using diamondanvil cells~\cite{diamondanvil-Rev2009-Bassett,diamondanvil-Rev2018-K.K.Mao}. In order to accomplish the metallization of hydrogen at relatively lower pressures, metal hydrides have been employed to utilize the so-called chemical precompression of hydrogen through metal elements~\cite{Ashcroft-MetalHydride}. Motivated by the theoretical predictions of high-$T_{\rm c}$ SC in a number of hydrides~\cite{LiHx-PANS2009,LiHx-acta cryst.2014,KHx-JPCC2012,CaH6-PANS2012,H3S-Sci.Rep2014,MgH6-RSC-Adv.2015,rare-earth-hydride-PRL2017,rare-earth-hydride-PANS2017}, experiments have been conducted to confirm that sulfur hydride H$_3$S exhibits a $T_{\rm c}$ of 203 K at pressures around 150 GPa~\cite{ExpH3S-Nature2015} and more recently, lanthanum hydride LaH$_{10}$ exhibits higher $T_{\rm c}$ around 250$-$260 K at ${\sim}$170 GPa~\cite{ExpLaH10-PRL2019, ExpLaH10-Nature2019}. Therefore, the combined theoretical and experimental breakthroughs in high-pressure compressed hydrides have triggered a new era of high-$T_{\rm c}$ superconductors~\cite{Hydride-Rev2020-Eremets,FeH5-Science2017,CeH9-Nat.Commun2019-T. Cui,CeH-Nat.Commun2019-J.F. Lin,Hydride-Rev2019-JCP}.

To search for metal hydrides with higher $T_{\rm c}$, there have been many theoretical studies of binary compounds including alkali metals~\cite{LiHx-PANS2009,LiHx-acta cryst.2014,KHx-JPCC2012}, alkaline earth metals~\cite{CaH6-PANS2012,MgH6-RSC-Adv.2015}, and rare earth metals~\cite{rare-earth-hydride-PRL2017,rare-earth-hydride-PANS2017,CeH9-Nat.Commun2019-T. Cui,CeH-Nat.Commun2019-J.F. Lin}. Recently, the realm of research has been extended to ternary compounds which may be more effective for achieving high-$T_{\rm c}$ SC because of an increase in the number of combinations of metal elements~\cite{Li2MgH16-PRL2019,CaYH12-PRL2019,LiPH6-Npj2019}. Based on first-principles density-functional theory (DFT) calculations and the Migdal-Eliashberg formalism, Yanming Ma and his colleagues predicted that a ternary hydride Li$_2$MgH$_{16}$ exhibits a $T_{\rm c}$ of ${\sim}$473 K at 250 GPa~\cite{Li2MgH16-PRL2019}. Such highest ever predicted $T_{\rm c}$ was enabled by Li doping in a binary hydride MgH$_{16}$ containing a large amount of H$_2$ molecules. Here, the supply of extra electrons via Li doping breaks the strong covalent bond of H$_2$ molecules to stabilize the clathrate H cages with weakly covalent H$-$H bonds. The resulting H network enhances the H-derived electronic density of states (DOS) at the Fermi level $E_{\rm F}$, giving rise to a high-$T_{\rm c}$ SC. It is natural that the charge transfer from Li dopants to H atoms in Li$_2$MgH$_{16}$ would be induced by a much lower electronegativity of Li atom compared to H atom~\cite{Li2MgH16-PRL2019}.

The concept of electronegativity describes the tendency of an atom to attract electron density towards itself. The electronegativity ${\chi}$ is usually assumed to be similar values in a variety of chemical environments. It is, however, noted that a recent quantum-mechanical model study~\cite{electronegativity-under-pressure} reported a significant variation of ${\chi}$ under pressure. Meanwhile, at high pressure, some materials behave as electrides, where some excess electrons are transferred from positively charged ions to interstitial regions~\cite{hoffman,hosono,Li6P}. Such loosely bound anionic electrons are here demonstrated to play an important role in the charge transfer process between Li dopants and H atoms in Li$_2$MgH$_{16}$.

\begin{figure}[ht]
\centering{ \includegraphics[width=8.0cm]{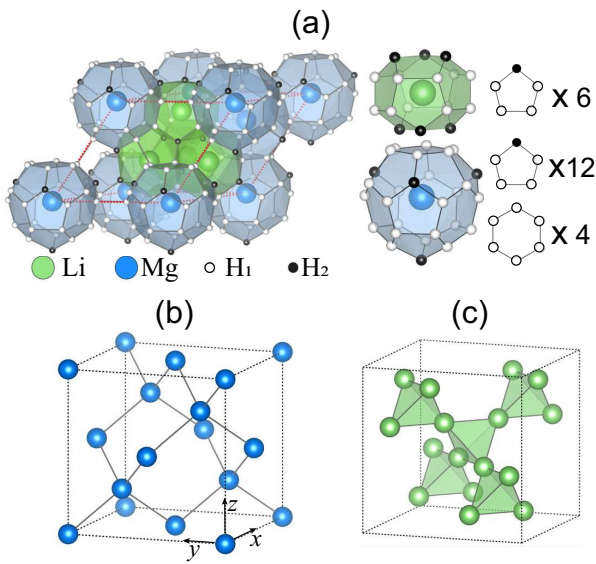} }
\caption{(Color online) (a) Optimized clathrate structure of Li$_2$MgH$_{16}$ consisting of Li-encapsulated H$_{18}$ cages and Mg-encapsulated H$_{28}$ cages. The H$_{18}$ cage consists of six pentagon rings, while the H$_{28}$ cage consists of twelve pentagon and four hexagon rings. There are two species of H atoms (designated as H$_{1}$ and H$_{2}$). The diamond lattice of Mg atoms and the pyroclore lattice of Li dopants are displayed in (b) and (c), respectively. The primitive unit cell with the lattice parameters $a$, $b$, and $c$ is shown in (a), while the conventional unit cell is shown in (b) and (c). In (b), the $x$, $y$, and $z$ axes point along the [001], [010], and [001] directions, respectively. }
\end{figure}

In the present study, we propose the underlying mechanism behind the charge transfer from Li dopants to H atoms in Li$_2$MgH$_{16}$. Using first-principles DFT calculations, we reveal that the Li dopants locating in the pyroclore lattice sites exhibit an electride feature with the anionic electrons residing in interstitial regions. Such loosely bound anionic electrons are easily captured to stabilize the clathrate H cages composed of two H species, H$_1$ and H$_2$ [see Fig. 1(a)]. It is thus likely that anionic electrons play an essential role in the charge transfer from Li dopants to H atoms. Our findings demonstrate that the anionic electrons created by the pyroclore-structured Li dopants are of vital importance not only for stabilizing the clathrate H cages, but also for enhancing the H-derived DOS at $E_{\rm F}$. The presently proposed charge transfer mechanism via anionic electrons can be also applied for compressed LaH$_{10}$~\cite{ExpLaH10-PRL2019, ExpLaH10-Nature2019}, and hence, it is anticipated to be more generic to other compressed high-$T_{\rm c}$ superconducting hydrides with clathrate structures.

\section{II. COMPUTATIONAL METHODS}

Our DFT calculations were performed using the Vienna {\it ab initio} simulation package with the projector-augmented wave method~\cite{vasp1,vasp2,paw}. Here, we included Li-1$s^2$2$s^1$, Mg-2$s^2$2$p^6$3$s^2$, and H-1$s^1$ electrons in the electronic-structure calculations. For the exchange-correlation energy, we employed the generalized-gradient approximation functional of Perdew-Burke-Ernzerhof (PBE)~\cite{pbe}. A plane-wave basis was taken with a kinetic energy cutoff of 850 eV. The ${\bf k}$-space integration was done with 16${\times}$16${\times}$16 $k$ points (in the Brillouin zone) for the structure optimization and 32${\times}$32${\times}$32 $k$ points for the DOS calculation. All atoms were allowed to relax along the calculated forces until all the residual force components were less than 0.001 eV/{\AA}. Using the QUANTUM ESPRESSO package~\cite{QE,oncv}, we calculated phonon frequencies with 4${\times}$4${\times}$4 $q$ points.

\begin{figure}[ht]
\centering{ \includegraphics[width=8.0cm]{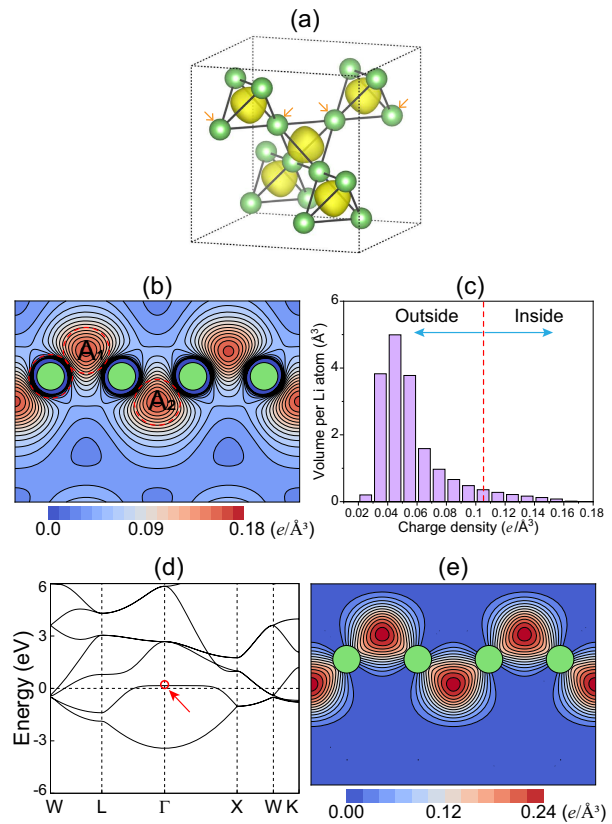} }
\caption{(Color online) (a) Isosurface and (b) contour plots of valence charge density of the isolated Li dopants with the pyroclore structure. The isosurface level in (a) is 0.12 $e$/{\AA}$^3$, and the contour spacing in (b) is 0.01 $e$/{\AA}$^3$. In (b), the contour plot is drawn on the (110) plane containing Li atoms indicated by the arrows in (a). The dashed circles in (b) represent the MT spheres around the Li atom with a radius of 0.713 {\AA} as well as around the A$_1$ and A$_2$ sites with a radius of 0.75 {\AA}. The histogram of volume distribution for every charge density range of 0.01 $e$/{\AA}$^3$ is given in (c), where ``Outside" and ``Inside" represent the charge density ranges outside and within the MT spheres around the A$_1$ and A$_2$ sites, respectively. The band structure of isolated Li dopants is displayed in (d). In (e), the charge density of the lowest unoccupied state at the ${\Gamma}$ point [indicated by the arrow in (d)] is plotted on the (110) plane with a contour spacing of 0.02 $e$/{\AA}$^3$.}
\end{figure}

\section{III. RESULTS}

We first optimize the geometry of Li$_2$MgH$_{16}$ at a pressure of 300 GPa~\cite{Li2MgH16-PRL2019}, which has a clathrate structure with the high crystalline symmetry of space group $Fd$$\overline{3}m$ [see Fig. 1(a)]. The optimized lattice parameters of the primitive unit cell are $a$ = $b$ = $c$ = 6.572 {\AA}. It is noted that Mg atoms form a diamond lattice [see Fig. 1(b)], while Li dopants form a pyroclore or three-dimensional (3D) kagome lattice [see Fig. 1(c)]. As shown in Fig. 1(a), there are two kinds of H cages: i.e., one is the H$_{18}$ cage surrounding a Li atom and the other is the H$_{28}$ cage surrounding a Mg atom. The H$_{18}$ cage consisting of six pentagon rings is opened to connect to neighboring H$_{18}$ cages, but the H$_{28}$ cage consisting of twelve pentagon and four hexagon rings has a closed shape [see Fig. 1(a) and Fig. S1 in the Supplemental Information~\cite{supple}]. The two cages are formed by two species of H atoms, H$_1$ (equivalent to H$_{\rm 96g}$ in Ref.~\cite{Li2MgH16-PRL2019}) and H$_2$ (equivalent to H$_{\rm 32e}$). We find that the H$_{1}-$H$_1$ bond length $d_1$ ($d_{1'}$) is 1.195 (1.023) {\AA}, while the H$_{1}-$H$_2$ bond length $d_2$ is 1.084 {\AA}. Note that the longer H$_{1}-$H$_1$ bond arises from the pentagon ring, while the shorter one is shared by the pentagon and hexagon rings (see Fig. S1 in the Supplemental Information~\cite{supple}). These values of $d_1$, $d_{1'}$, and $d_2$ are in good agreement with those ($d_1$ = 1.20, $d_{1'}$ = 1.02 {\AA}, and $d_2$ = 1.08 {\AA}) of a previous DFT calculation~\cite{Li2MgH16-PRL2019}.

Figure 2(a) shows the valence charge-density isosurface of the isolated Li dopants [see Fig. 1(c)] whose structure is taken from the optimized structure of Li$_2$MgH$_{16}$. Since the Li-1$s$ core state is located at around $-$46.6 eV below $E_{\rm F}$ [see Fig. S2(a) in the Supplemental Information~\cite{supple}], we exclude the 1$s^2$ core electrons to plot the valence charge density. It is noted that each Li atom in Fig. 2(b) has $\sim$0.08 electrons within the muffin-tin (MT) sphere having a radius of 0.713 {\AA} [close to the size of the corresponding Bader basin in Li$_{2}$MgH$_{16}$: see Fig. 3(d)], indicating that Li loses about 0.92 electrons. Interestingly, we find that some electrons detached from Li atoms are well distributed in the interstitial regions surrounded by four adjacent Li atoms [see Fig. 2(a)]. The confinement of such anionic electrons around the A$_1$ and A$_2$ sites is confirmed by the electron localization function~\cite{ELF} (see Fig. S3 in the Supplemental Material~\cite{supple}), which is effective for the characterization of interstitial electrons in electride materials~\cite{ELF-elect1,ELF-elect2,ELF-elect3}. It is, however, noticeable that anionic electrons arising from Li dopants are extensively distributed over the regions within and outside the MT spheres around the A$_1$ and A$_2$ sites [see Fig. 2(b)]. Figure 2(c) shows the histogram of volume distribution for every charge density range of 0.01$e$/{\AA}$^3$. This histogram reveals that, although anionic electrons have the highest densities at the A$_1$ and A$_2$ sites, they occupy more volume outside the MT spheres around the A1 and A2 sites compared to within the MT spheres. In order to examine how anionic electrons arising from Li dopants change as a function of pressure, we compare the valence charge densities of the isolated Li dopants at 250, 300, and 350 GPa, respectively (see
Fig. S4 in the Supplemental Information~\cite{supple}). We find that the charge density at the A$_1$ or A$_2$ site increases as 0.158, 0.164, and 0.170 $e$/{\AA}$^3$ at 250, 300, and 350 GPa, respectively, reflecting that anionic electrons around the A$_1$ and A$_2$ sites increase more dominantly with increasing pressure. It is thus likely that more anionic electrons can be captured to H cages with increasing pressure. Consequently, the H$_{1}-$H$_1$ and H$_{1}-$H$_2$ bond lengths are calculated to be shortened as ($d_1$, $d_{1'}$, $d_2$) = (1.229, 1.043, 1.106), (1.195, 1.023, 1.084), and (1.168, 1.006, 1.065) {\AA} at 250, 300, and 350 GPa, respectively. Based on these results, we can say that the Li dopants locating in the pyroclore lattice sites possess the electride characteristics with the A$_1$ and A$_2$ anions.

\begin{figure}[ht]
\centering{ \includegraphics[width=8.0cm]{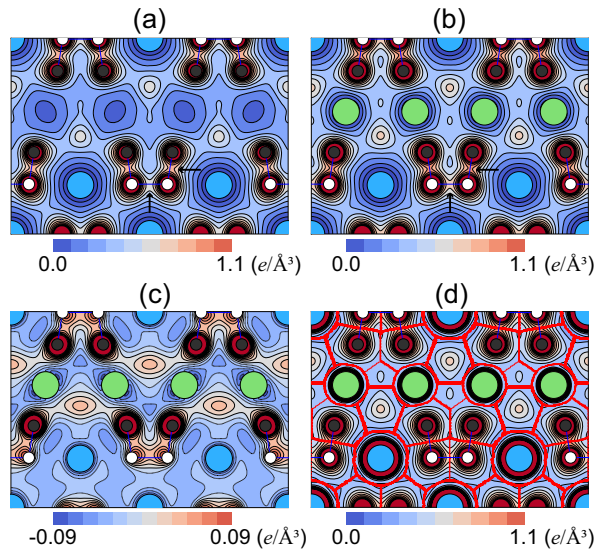} }
\caption{(Color online) Calculated valence charge densities of (a) the parent MgH$_{16}$ system and (b) Li$_2$MgH$_{16}$. The charge densities are drawn on the (110) plane with a contour spacing of 0.11 $e$/{\AA}$^3$. The charge density difference ${\Delta}{\rho}$ (defined in the text) is displayed in (c) with a contour spacing of 0.015 $e$/{\AA}$^3$. In (d), the Bader basins of Li$_2$MgH$_{16}$, obtained from the gradient of total charge density, is displayed. }
\end{figure}

In Fig. 2(d), we present the band structure of the pyroclore-structured Li dopants. It is seen that one band is fully filled but three bands are partially filled, with four valence electrons coming from four Li atoms in the primitive unit cell. Specifically, one band exhibits a partially flat dispersion at ${\sim}$0.1 eV above $E_{\rm F}$. The charge character of this flat band at the ${\Gamma}$ point represents strongly localized electrons at the interstitial regions surrounded by adjacent four Li atoms [see Fig. 2(e)], much larger than the total valence charge density shown in Fig. 2(b). Such a flatband nature of localized electrons is likely attributed to the geometric character of 3D kagome lattice, which hosts the destructive interferences of Bloch wave functions~\cite{Fe3Sn2-PRL}. The destructive interfered electrons producing dispersionless flatbands have been theoretically proposed in 2D kagome lattices~\cite{Kagome-JPA1992}, and their existence has been experimentally observed in real materials~\cite{Fe3Sn2-PRL}. We note that, despite the localized character of such interstitial electrons, the pyroclore-structured Li dopants show metallic behavior with multiple dispersive bands crossing $E_{\rm F}$ [see Fig. 2(c)].

\begin{figure*}[htb]
\centering{ \includegraphics[width=16.0cm]{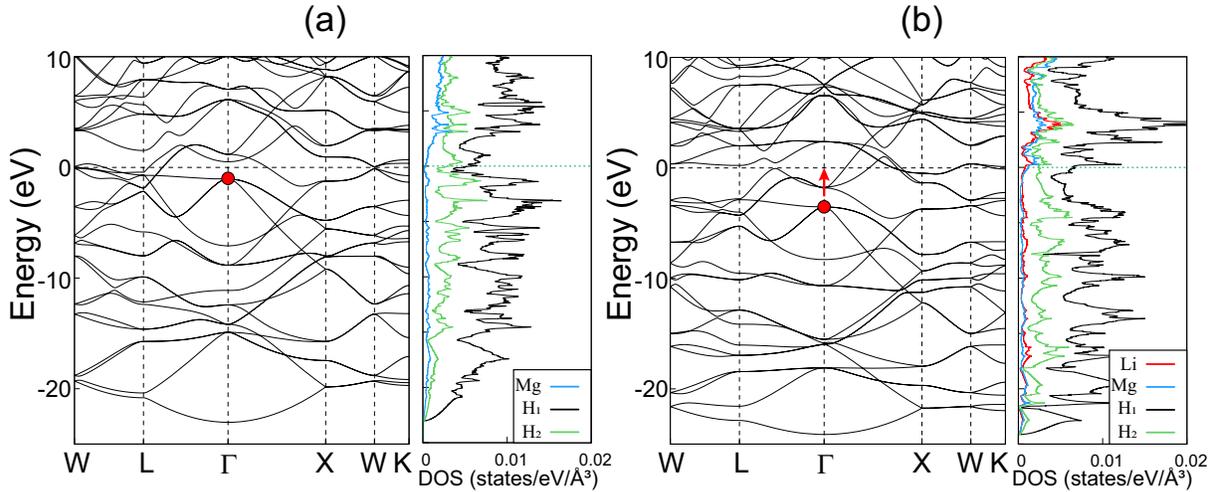} }
\caption{(Color online) Calculated band structures of (a) the parent MgH$_{16}$ system and (b) Li$_2$MgH$_{16}$, together with the corresponding LDOS of Mg/Li and H atoms. For the LDOS calculation, we choose the radii of muffin-tin spheres around Mg, Li, and H atoms as 0.94, 0.72 and 0.53 {\AA}, respectively. The energy zero represents $E_{\rm F}$. In (a), the highest occupied state at the ${\Gamma}$ point is marked with the red circle. In (b), the arrow indicates the shift of $E_{\rm F}$, relative to the parent MgH$_{16}$ system.}
\end{figure*}

Next, we explore the charge density of the parent MgH$_{16}$ system whose structure is taken from the optimized structure of Li$_2$MgH$_{16}$. Since the Mg-2$s$(2$p$) semicore states located at $-$77.8($-$44.6) eV below $E_{\rm F}$ are well separated from the valence states [see Fig. S2(b) in the Supplemental Information~\cite{supple}], we exclude such semicore electrons to plot the valence charge density of MgH$_{16}$ in Fig. 3(a). It is seen that H atoms in the H$_{18}$ and H$_{28}$ cages are bonded to each other with covalent bonds. We note that each H$-$H bond has a saddle point of charge density at its midpoint, similar to the C$-$C covalent bond in diamond~\cite{diamond}. The charge densities at the midpoints of the H$_{1}-$H$_1$ and H$_{1}-$H$_2$ bonds are 0.60 and 0.80 $e$/{\AA}$^3$, respectively [see the arrows in Fig. 3(a)]. It is, however, noticeable that the calculated phonon spectrum of MgH$_{16}$ exhibits imaginary frequencies in the whole Brillouin zone [see Fig. S5(a) in the Supplemental Material~\cite{supple}], indicating that the MgH$_{16}$ structure without Li atoms is dynamically unstable. We further relax the structure of MgH$_{16}$ to find a saddle point. The calculated phonon spectrum of such a saddle point structre also shows imaginary-frequency phonon modes [see Fig. S5(b)], indicating that it is still dynamically unstable. Meanwhile, the Li$_2$MgH$_{16}$ structure is dynamically stable without any imaginary-frequency mode [see Fig. S5(c) in the Supplemental Information~\cite{supple}]. It is thus likely that the stability of H cages in Li$_2$MgH$_{16}$ is enabled by capturing the anionic electrons in the pyroclore-structured Li dopants, as discussed below.

Figure 3(b) shows the calculated valence charge density ${\rho}_{\rm Li/MgH}$ of Li$_2$MgH$_{16}$, where the charge densities at the midpoints of the H$_{1}-$H$_1$ and H$_{1}-$H$_2$ bonds are 0.69 and 0.90 $e$/{\AA}$^3$, respectively [see the arrows in Fig. 3(b)]. These values are larger than the corresponding ones (0.60 and 0.80 $e$/{\AA}$^3$) in MgH$_{16}$, because the anionic electrons of the pyroclore-structured Li dopants are captured to the H$_{18}$ and H$_{28}$ cages. It is interestingly noticeable that the charge transfer from Li dopants to H atoms was interpreted in terms of a lower electronegativity of Li atom compared to H atom~\cite{Li2MgH16-PRL2019}. By contrast, we here propose that the electride nature of Li dopants is an essential ingredient in the charge transfer between Li dopants and H atoms, thereby providing new insight into understanding the charge transfer mechanism in Li$_2$MgH$_{16}$. In order to examine the charge transfer from the Li dopants to H cages, we calculate the charge density difference, defined as ${\Delta}{\rho}$ = ${\rho}_{\rm Li/MgH}$ $-$ ${\rho}_{\rm Li}$ $-$ ${\rho}_{\rm MgH}$, where ${\rho}_{\rm Li}$ and ${\rho}_{\rm MgH}$ represent the separated charge densities of Li dopants [see Fig. 2(b)] and parent MgH$_{16}$ [see Fig. 3(a)], respectively. As shown in Fig. 3(c), ${\Delta}{\rho}$ illustrates that electronic charge is transferred from Li [including the anion sites A$_1$ and A$_2$ in Fig. 2(b)] and Mg to H atoms. We further estimate the number of transferred electrons between the Li/Mg and H atoms by calculating the Bader charges~\cite{Bader} of Li$_2$MgH$_{16}$. Figure 3(d) shows the Bader basins of the constituent atoms, obtained from the gradient of the total charge density of Li$_2$MgH$_{16}$~\cite{Bader}. We find that the Bader charges inside Li, Mg, H$_1$, and H$_2$ basins are $-$2.22$e$, $-$8.34$e$, $-$1.18$e$, and $-$1.28$e$, respectively. Since this Bader charge of Li (Mg) includes 1$s^2$ (2$s^2$2$p^6$) core electrons, Li (Mg) atoms lose electrons of 0.78(1.66)$e$ per atom, while H$_1$ (H$_2$) atoms gain electrons of 0.18(0.28)$e$ per atom.

The charge transfer from Li dopants to H cages is expected to shift the Fermi level upward into the conduction band of the parent MgH$_{16}$ system. To explore the rigid band shift due to such an electron doping, we compare the band structures of MgH$_{16}$ and Li$_2$MgH$_{16}$, which are displayed in Figs. 4(a) and 4(b), respectively. The band dispersions of the two systems are generally similar to each other. As expected, we find that the empty conduction-band states of the MgH$_{16}$ system are occupied by Li doping, thereby leading to a shift of $E_{\rm F}$ by ${\sim}$2.49 eV [see the arrow in Fig. 4(b)]. It is noteworthy that the Li-derived states show their highest local DOS (LDOS) peak at around 3.9 eV above $E_{\rm F}$ [see Fig. 4(b)], indicating that Li$_2$MgH$_{16}$ is an electron-doped magnesium hydride via Li doping. Especially, the LDOS values of Li, Mg, H$_1$, and H$_2$ atoms at $E_{\rm F}$ are 0.8${\times}$10$^{-3}$, 1.2${\times}$10$^{-3}$, 2.5${\times}$10$^{-3}$, and 6.6${\times}$10$^{-3}$ states/eV/{\AA}$^3$, respectively [see Fig. 4(b)], indicating that the H-derived states are about 3$-$8 (2$-$5) times larger than Li (Mg)-derived states. Such large ratios of H-derived states in Li$_2$MgH$_{16}$ are different from the cases of other high-$T_{\rm c}$ hydrides such as H$_3$S~\cite{H3S-Sci.Rep2014} and LaH$_{10}$~\cite{LaH10-liangliang} where the LDOS of H-derived states at $E_{\rm F}$ is nearly the same as those of S- and La-derived states, respectively. Such large H-derived electronic states near $E_{\rm F}$ should increase electron-phonon coupling~\cite{Li2MgH16-PRL2019}, which leads to the highest $T_{\rm c}$ ever reported.

\section{IV. CONCLUSION}

Our first-principles DFT calculations have shown that the pyroclore-structured Li dopants exhibit an electride nature with anionic electrons ditributed in interstitial regions. Such loosely bound anionic electrons are easily captured to the H$_{18}$ and H$_{28}$ cages. We thus proposed that the electride nature of Li dopants importantly determines the charge transfer between Li dopants and H atoms. The present findings demonstrated that the anionic electrons created by Li dopants play important roles not only in stabilizing the clathrate H cages, but also in enhancing the H-derived DOS at $E_{\rm F}$. It is noted that the presently proposed charge transfer mechanism can be applicable to compressed LaH$_{10}$~\cite{LaH10-Seho} which was recently synthesized~\cite{ExpLaH10-PRL2019,ExpLaH10-Nature2019} to exhibit the highest $T_{\rm c}$ so far among experimentally available superconducting materials. Here, the charge transfer from La to H atoms is driven by the electride property of the La framework at high pressure (see Fig. S6 in the Supplemental Material~\cite{supple}). Therefore, the underlying mechanism of charge transfer in Li$_2$MgH$_{16}$ would be more generic, and it will be useful for the discovery and design of new high-$T_{\rm c}$ hydrides in the future.


\vspace{0.4cm}
\centerline{\bf ACKNOWLEDGEMENTS}
\vspace{0.4cm}

This work was supported by the National Research Foundation of Korea (NRF) grant funded by the Korean Government (Grants No. 2019R1A2C1002975, No. 2016K1A4A3914691, and No. 2015M3D1A1070609). The calculations were performed by the KISTI Supercomputing Center through the Strategic Support Program (Program No. KSC-2019-CRE-0183) for the supercomputing application research.  \\

\noindent $^{*}$ Corresponding author: chojh@hanyang.ac.kr

\newpage
\onecolumngrid
\newpage
\titleformat*{\section}{\LARGE\bfseries}

\renewcommand{\thefigure}{S\arabic{figure}}
\setcounter{figure}{0}

\vspace{1.2cm}

\section{Supplemental Material for “Underlying mechanism of charge transfer in Li-doped MgH$_{16}$ at high pressure”}
\vspace{1.8cm}
\begin{flushleft}

{\bf 1. H-H bond lengths of the H$_{18}$ and H$_{28}$ cages }
\begin{figure}[ht]
\includegraphics[width=11.5cm]{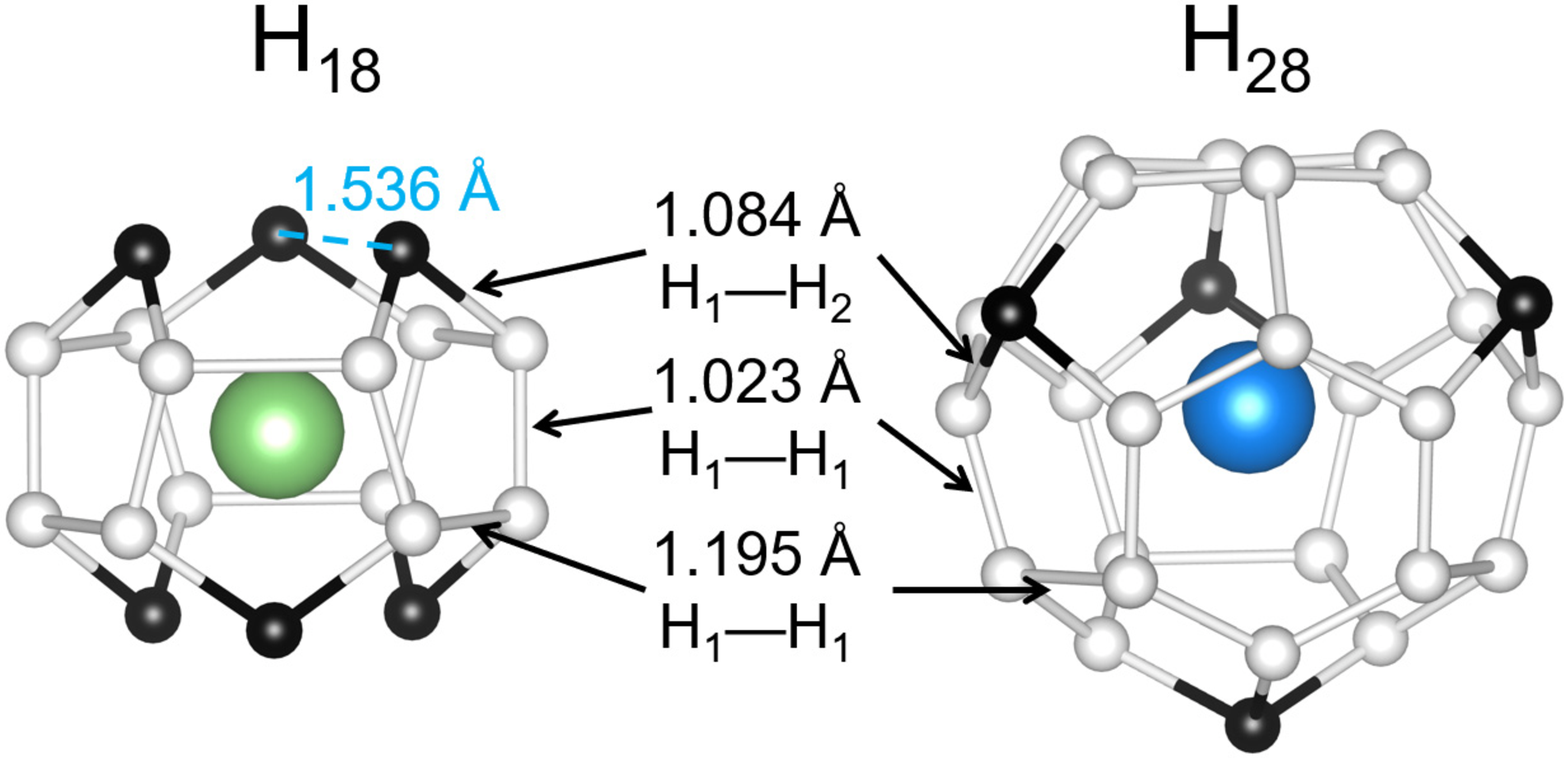}
\caption{ Calculated H-H bond lengths of the H$_{18}$ and H$_{28}$ cages at 300 GPa. The H$_{18}$ cage consisting of six pentagon rings is opened, but the H$_{28}$ cage consisting of twelve pentagon and four hexagon rings has a closed shape. }
\end{figure}

\vspace{2cm}

{\bf 2. Band structures of the pyroclore-structured Li dopants, MgH$_{16}$, and Li$_2$MgH$_{16}$ }
\begin{figure}[ht]
\includegraphics[width=17.5cm]{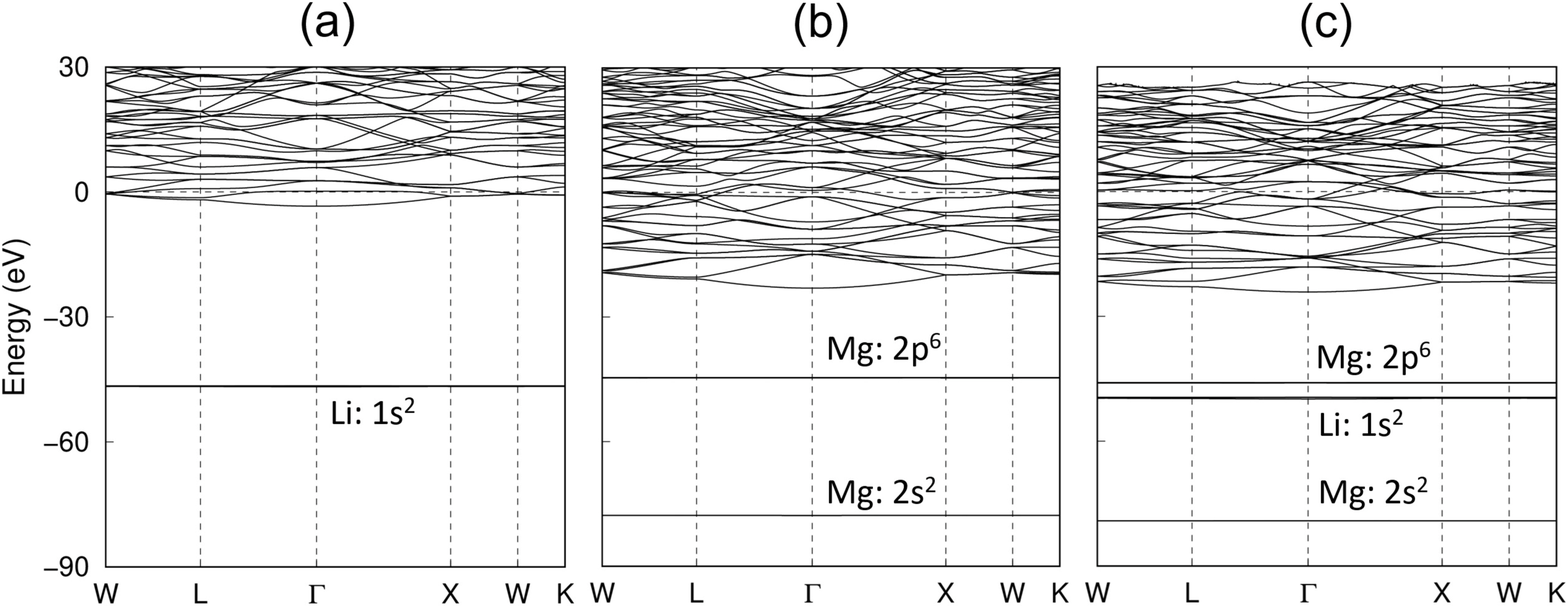}
\caption{ Calculated band structures of (a) the pyroclore-structured Li dopants, (b) MgH$_{16}$, and (c) Li$_2$MgH$_{16}$. The electronic states of Li-1$s$$^2$ or/and Mg-2$s$$^2$2$p$$^6$ are well separated from the valence states of each system. }
\end{figure}

\newpage

{\bf 3. Electron localization function of the pyroclore-structured Li dopants}
\begin{figure}[ht]
\includegraphics[width=7.5cm]{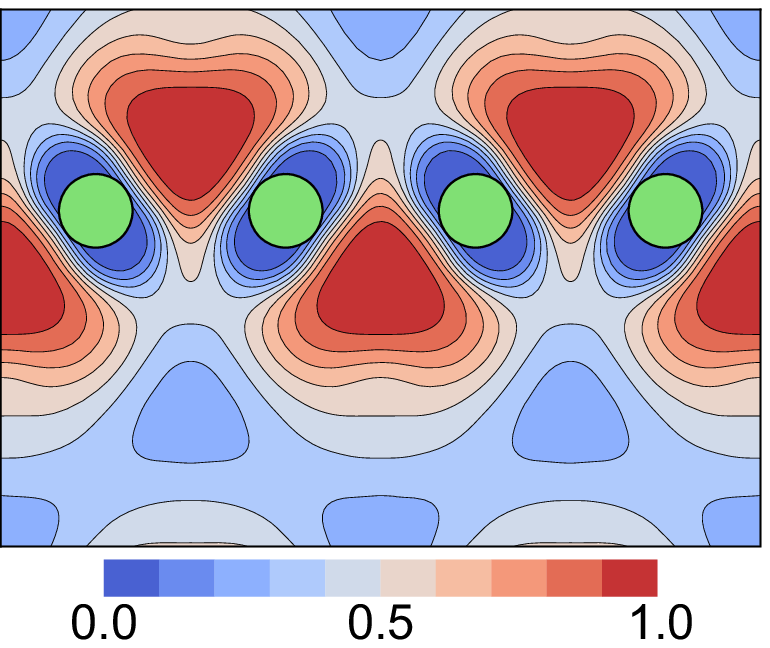}
\caption{ Electron localization function of the pyroclore-structured Li dopants. This result indicates that some electrons are transferred from Li atoms to the interstitial regions around the A$_1$ and A$_2$ sites [see Fig. 2(b)]. }
\end{figure}

\vspace{3.2cm}

{\bf 4. Valence charge densities of the isolated Li dopants at 250, 300, and 350 GPa, respectively}
\begin{figure}[ht]
\includegraphics[width=17.5cm]{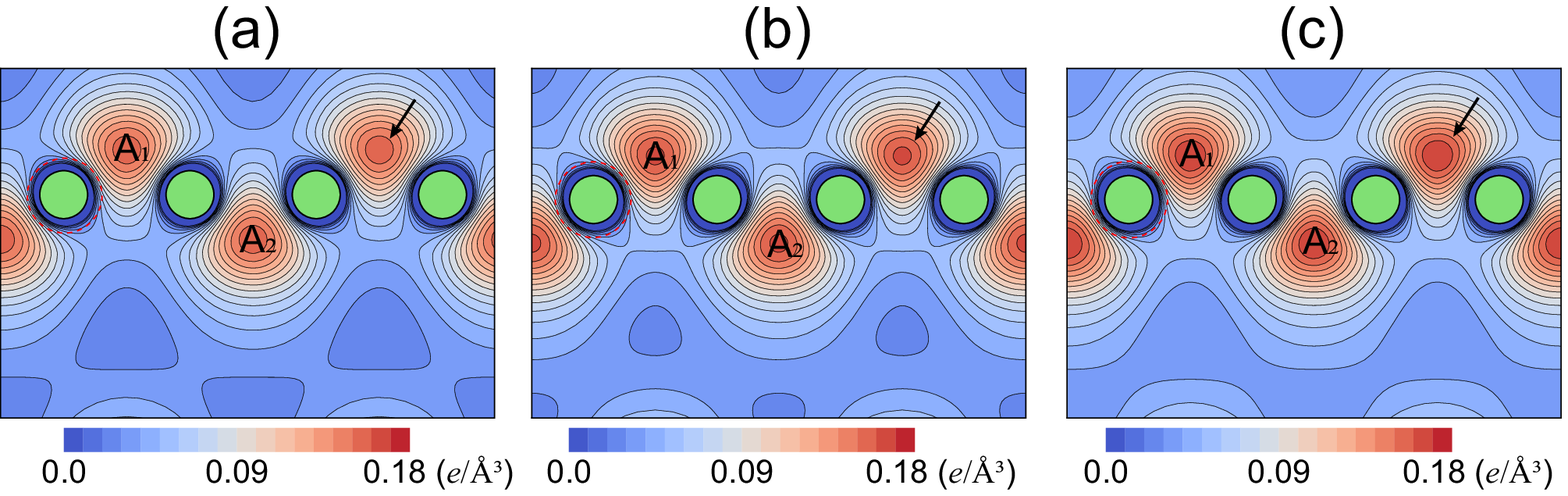}
\caption{ Contour plots of valence charge density of the isolated Li dopants at (a) 250, (b) 300, and (c) 350 GPa. These plots are drawn on the (110) plane with a contour spacing of 0.01 $e$/{\AA}$^3$. The black arrow indicates the charge density contour line of 0.15 $e$/{\AA}$^3$. }
\end{figure}

\newpage

{\bf 5. Phonon spectra of MgH$_{16}$ and Li$_2$MgH$_{16}$}
\begin{figure}[ht]
\includegraphics[width=16.5cm]{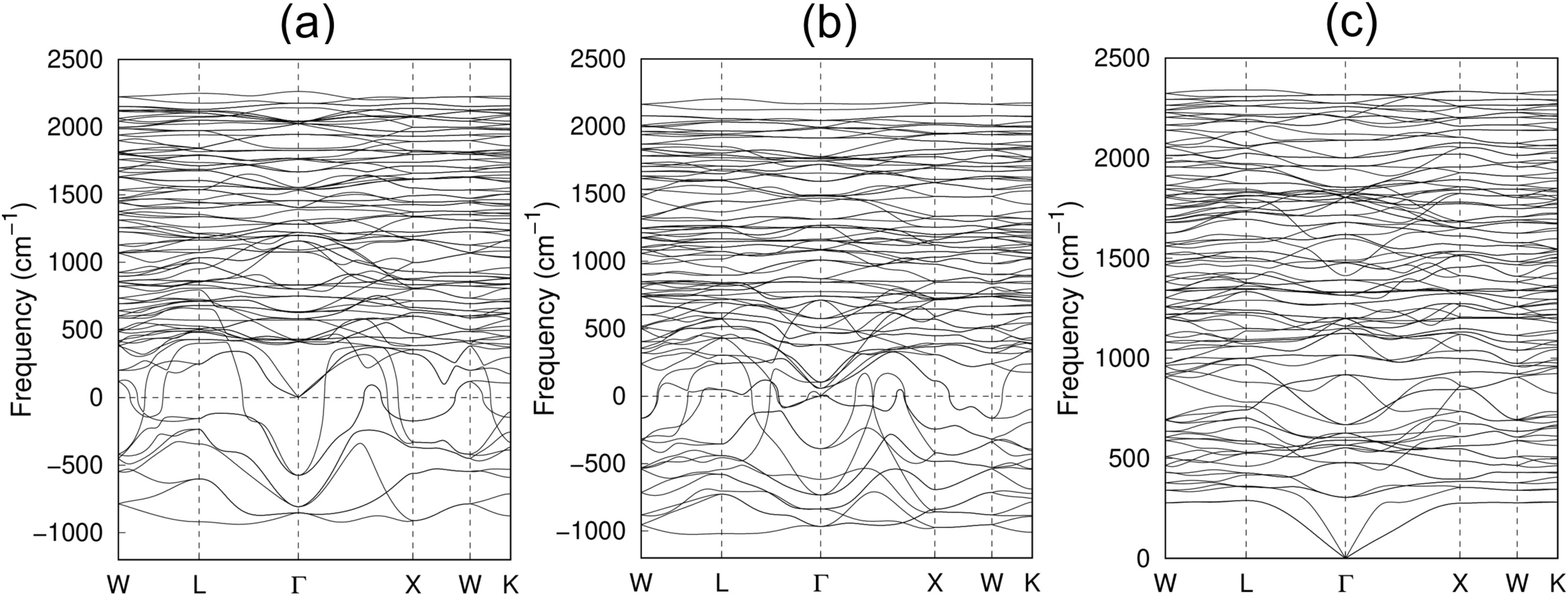}
\caption{ Calculated phonon spectra of (a) MgH$_{16}$, (b) relaxed MgH$_{16}$, and (c) Li$_2$MgH$_{16}$. We find that MgH$_{16}$ without Li atoms is dynamically unstable, while Li$_2$MgH$_{16}$ is dynamically stable. }
\end{figure}

\vspace{1.2cm}

{\bf 6. Structure of LaH$_{10}$ and anionic electrons in the La lattice}
\begin{figure}[ht]
\includegraphics[width=9.5cm]{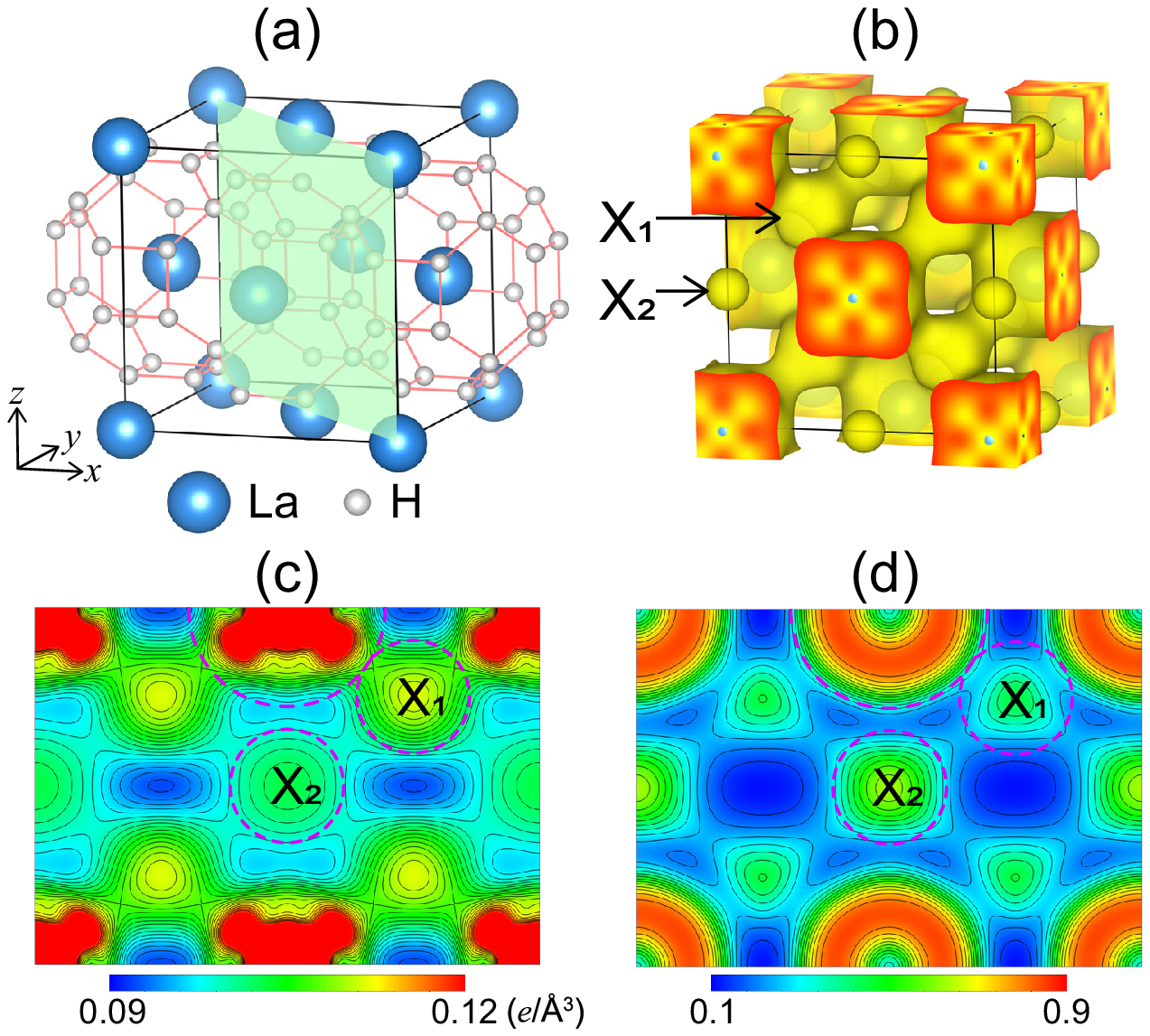}
\caption{ Optimized structure of LaH$_{10}$ at 300 GPa and (b) calculated charge density isosurface of valence electrons in the La lattice whose structure is taken from LaH$_{10}$. The charge density and electron localization function (ELF) of the La lattice are plotted on the (110) plane in (c) and (d), respectively. A (110) plane is drawn in (a). The isosurface in (b) is taken as 0.11 $e$/{\AA}$^3$, and the contour spacings in (c) and (d) are 0.002 $e$/{\AA}$^3$ and 0.05, respectively. In (b), X$_1$ and X$_2$ indicate the two anions in interstitial regions. The dashed circles in (c) represent the muffin-tin spheres around La, X$_1$ and X$_2$ with the radii of 1.31, 0.75 and 0.75 /{\AA}, whithin which the electron numbers are 1.17$e$, 0.21$e$ and 0.19$e$, respectively. Therefore, each La atom loses about 1.83 electrons, some of which are transferred to the interstitial regions around the X$_1$ and X$_2$ sites. The confinement of anionic electrons around the X$_1$ and X$_2$ sites is confirmed by the ELF [see T. Tada et al., Inorg. Chem. 53, 10347 (2014)]. It is thus likely that the charge transfer from La to H atoms in compressed LaH$_{10}$ is driven by the electride property of the La lattice. }
\end{figure}

\end{flushleft}

\end{document}